# Early evolution of purple retinal pigments on Earth and implications for exoplanet biosignatures

Shiladitya DasSarma[1] and Edward W. Schwieterman[2,3,4,5]

[1]Department of Microbiology and Immunology, University of Maryland School of Medicine, Institute of Marine and Environmental Technology, Baltimore, MD, USA; [2]Department of Earth Sciences, University of California, Riverside, CA, USA; [3]NASA Postdoctoral Program Fellow, Universities Space Research Association, Columbia, MD, USA; [4]NASA Astrobiology Institute's Alternative Earths and Virtual Planetary Laboratory Teams and [5]Blue Marble Space Institute of Science, Seattle, WA, USA

## Abstract

We propose that retinal-based phototrophy arose early in the evolution of life on Earth, profoundly impacting the development of photosynthesis and creating implications for the search for life beyond our planet. While the early evolutionary history of phototrophy is largely in the realm of the unknown, the onset of oxygenic photosynthesis in primitive cyanobacteria significantly altered the Earth's atmosphere by contributing to the rise of oxygen ∼2.3 billion years ago. However, photosynthetic chlorophyll and bacterio chlorophyll pigments lack appreciable absorption at wavelengths about 500–600 nm, an energy-rich region of the solar spectrum. By contrast, simpler retinal-based light-harvesting systems such as the haloarchaeal purple membrane protein bacteriorhodopsin show a strong well-defined peak of absorbance centred at 568 nm, which is complementary to that of chlorophyll pigments. We propose a scenario where simple retinal-based light-harvesting systems like that of the purple chromoprotein bacteriorhodopsin, originally discovered in halophilic Archaea, may have dominated prior to the development of photosynthesis. We explore this hypothesis, termed the 'Purple Earth,' and discuss how retinal photopigments may serve as remote biosignatures for exoplanet research.

## Background

The major events sparking life on Earth on our 4.6-billion-year-old planet remain enigmatic, although there is general agreement that first life likely arose about 3.7–4.1 billion years ago, during the early Archean or late Hadean eons (Abramov and Mojzsis, 2009; Deamer, 2011; Bell *et al*., 2015; Knoll, 2015). Evidence for the presence of isoprenoid compounds has been reported in ancient sediments not long after, suggesting the early rise of Archaea (Hahn and Haug, 1986; Ventura *et al*., 2007). The early rise of Archaea is also suggested by phylogenic studies, although lateral gene transfers have complicated their interpretation (Lange *et al*., 2000; Kennedy *et al*., 2001; Brochier-Armanet *et al*., 2011; Hoshino and Gaucher, 2018). Stromatolites representing fossilized microbial mats have been estimated to be up to 3.7 billion-years-old (Walter *et al*., 1980; Vankranendonk *et al*., 2008; Nutman *et al*., 2016) and radiocarbon dating has shown $^{12}$C enrichment from this early period, consistent with the development of photosynthetic microorganisms (Ohtomo *et al*., 2014). There is wide agreement that anoxygenic photosynthesis preceded oxygenic photosynthesis, though the length of the interval for this transition is uncertain (Olson, 2006; Buick, 2008; Rothschild, 2008). Some geochemical proxy records suggest that the earliest oxygenic photosynthesizers may have appeared by ∼2.9–3 Ga with geochemical sinks arresting oxygen's accumulation for a time (Nisbet *et al*., 2007; Planavsky *et al*., 2014). Ultimately, because of oxygenic photosynthesis and additional, poorly understood factors, the Earth experienced a Great Oxidation Event about 2.3 billion years ago, which indelibly altered the prevailing chemical conditions of our planet's atmosphere (Kump, 2008; Lyons *et al*., 2014; Luo *et al*., 2016).

What were the important evolutionary events predating the rise of photosynthesis during the early history of life on Earth? Although the events during this very early time are not clear, in this paper, we discuss a speculative hypothesis for early evolution, called the 'Purple Earth,' which posits the rise of retinal pigment-based phototrophic life forms on Earth's surface prior to anoxygenic and oxygenic photosynthesis. In this view, retinal pigments may have competed with and affected the evolution of photosynthetic pigments and indeed still complements them today in Earth's oceans and other environments. Early microorganisms employing retinal pigments for generating metabolic energy may have dominated, as halophilic Archaea do today in hypersaline environments, providing a scenario which may serve to guide our search for detectable biosignatures on other worlds.

### Early evolution on Earth

During the first half of Earth's history, stretching over 2 billion years, dramatic and long-lasting evolutionary inventions occurred through processes that we are only beginning to understand (Fig. 1; Deamer, 2011; Knoll, 2015). They include prebiotic evolution and the development of cellularity, the foundation of the last universal common ancestor (LUCA) and evolution of the universal genetic code (Fenchel, 2002). Other factors critical for the success of early life were the evolution of transmembrane potential and chemiosmotic coupling for creating and storing bioenergy, pigments for the capture of light energy for phototrophy and photosynthesis and respiratory chains for anaerobic and aerobic respiration (Zannoni, 2004). In addition, a 'frozen accident' has been proposed to establish the genetic code as a universal feature in all extant life on Earth (Crick, 1968; Söll and RajBhandary, 2006). During the earliest period in evolutionary history, well-defined phylogenetic lineages may not yet have been established; instead extensive lateral gene transfers allowed for ready sharing of new innovations until such time when the last common ancestor experienced competitive selective forces and diverged into the three primary 'Domains of Life' (Woese, 2002).

Even prior to the evolution of the three Domains, the development of a protocell must have been facilitated by the evolution of a water-tight cell membrane as a permeability barrier, preventing the free diffusion of chemicals into and out of cells, critical for generating and storing cellular energy (Gunner et al., 2013). The intracellular milieu provided a microenvironment in which biomolecular functions, such as the biosynthesis of macromolecules and the genetic code could be established. Transmembrane ion pumps acting as energy transduction and storage systems must have been among the earliest inventions. In one scenario proposed here, a simple light-harvesting system incorporating a retinal pigment allowed light-driven proton pumping and led to a proton-motive gradient. Based on its ubiquity, the transmembrane electrochemical potential (i.e. proton-motive gradient) as well as phosphoric anhydride bonds, such as in adenosine triphosphate (ATP), became established and universal due to their kinetic stability and bioenergetic capabilities in the aqueous environment. Subsequently, retinal as well as a variety of more complex anaerobic and oxygenic light-harvesting systems were invented and resulted in the evolution of diverse phototrophic and photosynthetic microorganisms.

### Appearance of purple retinal pigments

The earliest life-forms probably arose in the early Archean or possibly late Hadean Eons, with some molecular clock estimates putting life's origin as early as 4 Ga (Hedges, 2002). While the exact timing of appearance of retinal pigments is not clear, it may have been a very early metabolic invention coincident with or occurring soon after the development of cellular life. A retinal chromophore bound to a single polypeptide allows a system for phototrophy by forming a chromoprotein, like bacteriorhodopsin in halophilic Archaea dominant in hypersaline environments and proteorhodopsin in pelagic bacteria distributed throughout the oceans (Béjà et al., 2001; Stoeckenius et al., 1979). The absorption of light by this chromoprotein in the 490–600 nm region, a highly energy-rich region of the solar spectrum (Fig. 2), is directly coupled to pumping of protons and the resulting electrochemical gradient chemiosmotically drives ATP synthesis. This type of retinal-dependent phototrophy is considerably simpler albeit less efficient than photosynthesis and it neither results in fixation of carbon nor production of oxygen (Pinhassi et al., 2016). Nevertheless, the widespread distribution of retinal chromoproteins in nature and their unique utilization of the energy-rich, yellow-green region of the spectrum for production of cellular energy suggest their early appearance on Earth.

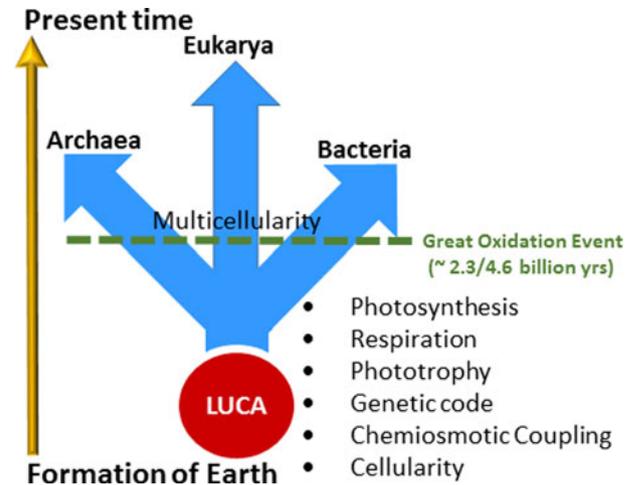

**Fig. 1.** Evolutionary timeline and events. The arrow at the left roughly indicates time from the formation of the Earth to the present, about 4.6 billion years. Geochemical and fossil evidence indicate that life arose soon after the Earth formed, with many key evolutionary inventions following: cellularity, chemiosmotic coupling, genetic code, phototrophy, respiration and photosynthesis. Light-driven proton pumping by retinal proteins are hypothesized to have evolved during this early stage in evolution. The last universal common ancestor (LUCA) predated the divergence of life into three Domains: Archaea, Bacteria and Eukarya. The rise of anoxygenic and then oxygenic photosynthesis allowed the productivity of Earth's microbial biosphere to increase immensely (Des Marais, 2000). The Great Oxidation Event followed, about 2.3 billion years ago and led to the development of multicellularity and evolution of higher life forms.

Evidence for the existence of isoprenoid compounds that are part of the biosynthetic pathway to retinal as well as archaeal lipids in the early history of the Earth has also been provided (Hahn and Haug, 1986; Ventura et al., 2007). It is likely that the evolutionary invention of retinal pigments was coincident with other membrane lipids, which together established the molecular basis for chemiosmotic coupling and phototrophic capabilities (Boucher and Doolittle, 2000). Retinal is produced by a branch of the isoprenoid metabolic pathway leading to carotenoids and branched-chain lipids, which are found in cell membranes (Fig. 3). Retinal pigments occur in both major prokaryotic phylogenetic groups, Archaea and Bacteria, as well as in eukaryotes, where they are essential components of the visual system (Ernst et al., 2014). Among the pigments prevalent in nature, retinal has a simple structure compared with many others that are used for photosynthesis and respiration, e.g. chlorophyll and other porphyrins, which may be produced by a branch of the tricarboxylic acid (TCA) cycle, a pathway used by all aerobic organisms (Mailloux et al., 2007). These findings, together with the central position of retinal at the intersection of lipid metabolism and bioenergetics, as well as its widespread distribution suggest that retinal played an important role in the early evolution of life on Earth.

The light-driven proton pumping activity of retinal pigments such as the chromoprotein bacteriorhodopsin in the membrane of an early cell would have allowed the development of chemiosmotic coupling, linking of membrane potential to other transmembrane transport processes and ATP synthesis (Stoeckenius

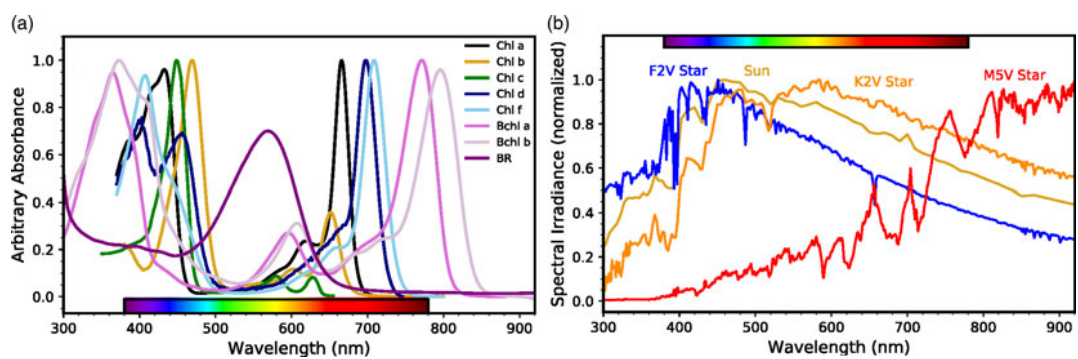

**Fig. 2.** Phototrophic pigment absorption and stellar radiation as a function of wavelength. (a) Absorbance spectra of phototrophic pigments including chlorophyll a, b, c, d and f (Chen *et al.*, 2010; Chen and Blankenship, 2011; Jeffrey, 1963); bacteriochlorophyll a and b (Frigaard *et al.*, 1996); and bacteriorhodopsin (BR; credit: Victoria Laye and Priya DasSarma). Note the strong BR absorption where (bacterio)chlorophylls are least absorptive. (b) Normalized spectral energy distributions at the top of the atmosphere for FGKM-type stars, including the Sun (G-type), from the Virtual Planetary Laboratory (Meadows *et al.*, 2018; Segura *et al.*, 2003).

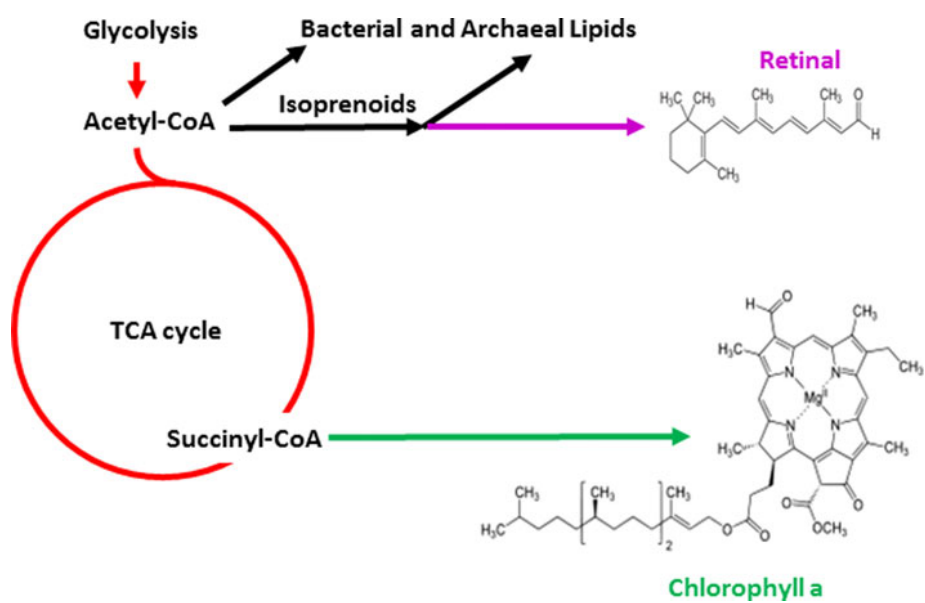

**Fig. 3.** Biosynthetic pathways for photopigments. Pathways leading to retinal (purple) and chlorophyll (green) branching from central metabolism (red) are shown. Glycolysis and the TCA cycle are depicted as are structures of the simpler retinal chromophore and the more complex chlorophyll a.

*et al.*, 1979). A retinal-based phototrophic system clearly represents one of the simplest bioenergetic mechanisms conceivable, requiring only a single opsin inserted in a membrane vesicle and membrane-potential coupled ATP synthase (Fig. 4). Indeed, such a model phototrophic system, inside out, was established *in vitro* in the 1970s using haloarchaeal bacteriorhodopsin and mitochondrial ATP synthase in artificial lipid vesicles (Racker and Stoeckenius, 1974). This seminal work was credited with helping to establish the validity of Mitchell's chemiosmotic coupling hypothesis (Mitchell, 1961) and also forms the foundation of one of the simplest and as proposed, earliest metabolic capabilities in evolution, retinal-based phototrophy.

Early Earth environments would have lacked abundant free $O_2$ in contrast to highly oxic modern environments and required the production of retinal using a terminal oxidative step in a likely strictly anaerobic environment. A number of potential mechanisms have been proposed for generating such an oxidative potential, such as pyrite-induced aqueous hydrogen peroxide and hydroxide radical formation (Borda *et al.*, 2001; Cohn *et al.*, 2006). Other anaerobic oxidation reactions are also known, such as anaerobic oxidation of methane and ammonium and transformation of isoprenoids by anaerobic microorganisms (Hallam *et al.*, 2004; Hylemon and Harder, 1998; Strous and Jetten, 2004).

Also notable is that modern halophilic Archaea are facultative, rather than obligate aerobes and can respire nitrate and TMAO/DMSO (Mancinelli and Hochstein, 1986; Müller and DasSarma, 2005). Indeed, Haloarchaea have been shown to engage in phototrophy in microaerobic or anoxic laboratory conditions (Sumper *et al.*, 1976; DasSarma *et al.*, 2012; Laye *et al.*, 2017). Additionally, a considerable amount of evidence suggests that the genes for aerobic respiration were laterally transferred to halophilic Archaea (Kennedy *et al.*, 2001) and their ultimate origin may have been as an anaerobic chemolithoautotrophic methanogen (Nelson-Sathi *et al.*, 2012; Aouad *et al.*, 2018). Hence, haloarchaeal phototrophic metabolism was probably developed well before genes for aerobic respiration were acquired, possibly in Archaea inhabiting hypersaline environments (Stevenson *et al.*, 2015). While these modern haloarchaeal organisms have certainly changed over the eons from the original retinal-based phototrophs, the available evidence illustrates the potential capacity of Haloarchaea to have survived the anaerobic conditions that prevailed on ancient Earth.

In modern halophilic Archaea, the retinal protein bacteriorhodopsin trimers form a hexagonal lattice which can cover a large fraction of the cell surface (Stoeckenius *et al.*, 1979), imparting a bright purple colour to some salt ponds where they dominate (Fig. 5). The resulting purple membrane can be easily isolated using sucrose density gradients and has been the subject of extensive structural and functional analysis of transmembrane ion translocation (Henderson and Unwin, 1975; Stoeckenius *et al.*, 1979; Krebs and Khorana, 1993; Hirai *et al.*, 2009). Bacteriorhodopsin is a prototype of integral membrane proteins with seven-transmembrane α-helical segments where the retinal chromophore is bound by a Schiff's base linkage to the ϵ-amino group of a lysine residue (Bayley *et al.*, 1981). The photobiology of bacteriorhodopsin has been intensively studied, including characterization of the molecular dynamics and role of retinal during photocycling (Hirai *et al.*, 2009). The bacteriorhodopsin resting state is notable for the characteristic colour purple resulting from the strong absorption peak maximum at 568 nm in the yellow-green region of the spectrum.

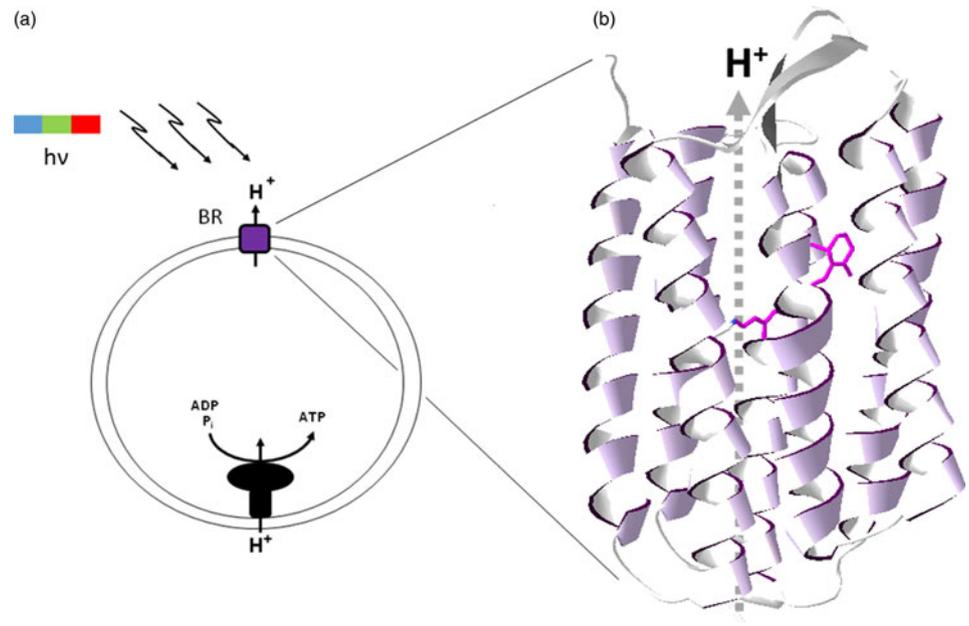

**Fig. 4.** Bacteriorhodopsin and chemiosmotic coupling. (a) Light-driven (hν) proton pumping by bacteriorhodopsin (BR) results in ATP synthesis by chemiosmotically coupling to the proton-motive force. (b) Bacteriorhodopsin structure showing seven-transmembrane α-helical segments (ribbons) and bound retinal chromophore (purple wire structure), with proton pumping (dashed arrow, H⁺).

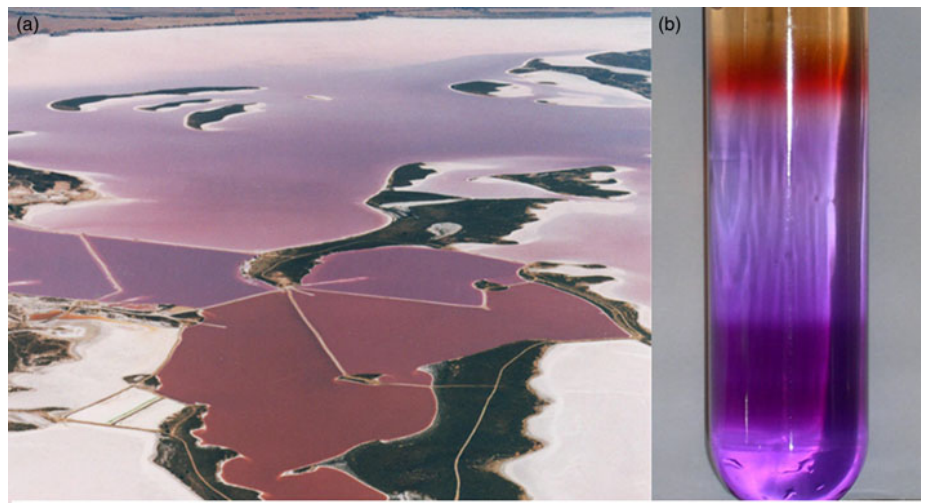

**Fig. 5.** Purple microorganisms and purple membrane. (a) Australian salt pond with a bloom of purple microorganisms (Courtesy Cheetham Salt Co.). (b) Sucrose gradient separating *Halobacterium* sp. cell lysate, including both red (upper) and purple (lower) pigments (Credit: Victoria Laye and Priya DasSarma).

### Spectral complementarity of photopigments

Comparison of the spectrum of bacteriorhodopsin with the major photosynthetic pigments containing chlorophyll and bacteriochlorophylls shows them to be complementary, i.e. the purple pigment absorption peaks in the region with a trough for the green pigments (Fig. 2a). If the evolution of the simpler retinal pigments predated chlorophyll pigments in the evolutionary history, as proposed, it is conceivable that they may have affected the development of the spectral characteristics of evolving chlorophyll pigments (Goldsworthy, 1987; DasSarma, 2006). This may have been the consequence of filtering of light by retinal chromoproteins, resulting in a deficit of wavelengths of light centred around the peak of bacteriorhodopsin absorption in the yellow-green region of the spectrum. The resulting deficit, particularly in a stratified community of microorganisms such as those observed in stromatolites, may explain why chlorophylls and bacteriochlorophylls evolved to absorb relatively little light in the yellow-green energy-rich portion of the electromagnetic spectrum, instead absorbing light primarily in the flanking blue and red regions of the solar spectrum.

Modern stromatolites represent microbial communities with on-going spectral competition and spectral tuning of chromoproteins (Croce and van Amerongen, 2014). Phototrophic and photosynthetic microorganisms in microbial mats are commonly stratified based on predictable photosystem characteristics as well as oxygen requirements. In such communities, oxygenic cyanobacteria are found near the surface oxic zone while anaerobic phototrophic and photosynthetic microbes are buried at lower anoxic regions. If these modern stratified microbial communities are like those present in ancient stromatolites, filtering of wavelengths of light would have been an important and pervasive characteristic of microbial communities. Modern microbial communities all over the world support planktonic retinal-containing halophilic Archaea and Bacteria inhabiting brines above photosynthetic mats (Cohen and Rosenberg, 1989). If co-evolution of retinal and chlorophyll photopigments occurred in deep evolutionary history, stratification within such niches may have played an important role in the evolution of spectral properties.

Importantly, modern rhodopsin-based phototrophy is present throughout the oceanic and terrestrial biosphere, including non-hypersaline conditions and environments that may have been common in the distant geologic past. While originally discovered in halophilic Archaea, microbial rhodopsins are common in oceanic planktonic Bacteria (Kandori, 2015). For example, *Pelagibacter ubique* is a widely distributed marine bacterium that produces the retinal chromoprotein, proteorhodopsin, with the ability to use its light-driven proton pumping activity for energy generation. Moreover, the absorption characteristics of proteorhodopsins show spectral variations in oceanic planktonic bacteria isolated from different depths, consistent with spectral tuning (Rangarajan *et al.*, 2007). Rhodopsin-based phototrophy in the ocean may be so widespread as to rival the total light capture of photosynthesizers (Brown, 2014; Gómez-Consarnau *et al.*, 2017). Furthermore, metagenomic analyses have recently uncovered evidence for the widespread presence of rhodopsins in the terrestrial biosphere including in the phyllosphere (leaf surfaces) and even edaphic systems and hypolithic communities in the Antarctic Dry Valley (Atamna-Ismaeel *et al.*, 2012; Guerrero *et al.*, 2017). These findings are consistent with the notion that microorganisms evolve specialized photosystems that make use of any available spectral region with sufficient energy. The widespread presence of microbial rhodopsins in modern environments, along with inefficient chlorophyll absorption in the middle of the visible spectrum where rhodopsin light capture is most efficient, suggests a co-evolution that is consistent with the earlier appearance of retinal-based phototrophy (PBS Eons 2018).

### Rise of photosynthesis

The rise of anaerobic and oxygenic photosynthesis and retreat of retinal-based life must have occurred in discrete stages, which are not fully understood. For example, the time of appearance of a wide diversity of bacteria with anoxygenic photosynthetic systems is not precisely known (Jeffrey, 1963; Frigaard *et al.*, 1996; Rothschild, 2008; Chen *et al.*, 2010; Chen and Blankenship, 2011; Croce and van Amerongen, 2014). The purple and green bacteria possessing bacteriochlorophyll may have been an early evolutionary development with photosynthetic reaction centres evolving from electron transport chain components such as cytochromes (Williamson *et al.*, 2011; Mazor *et al.*, 2012). Alternatively, a simplified photosystem may have evolved previously, like those in some heliobacteria (Xiong *et al.*, 1998). In either case, evolution likely first led to the evolution of anoxygenic photosynthesis with the development of more complex oxygenic photosynthetic membrane systems like those in modern cyanobacteria, including two photosynthetic reaction centres and a host of membrane components, developing later.

With multiple evolutionary steps leading to progressively higher efficiency chlorophyll pigments along with the invention of accessory pigments, photosynthetic microorganisms out-competed retinal-based phototrophic microorganisms in most environments. Evolution of anoxygenic photosynthesizers was followed by oxygenic photosynthesizing cyanobacteria and ultimately eukaryotic algae and plants. Interestingly, a distinct hypothesis that purple sulphur bacteria may have dominated euxinic oceans during the mid-Proterozoic eon has also been proposed, resulting in a *second* Purple Earth (Brocks *et al.*, 2005; Sanromá *et al.*, 2014), which would have been long after the retreat of retinal-based life dominating the *first* Purple Earth to ecological niches resembling those of today. The development of eukaryotic algae and complex plants and their spread throughout the terrestrial environment allowed the evolution of land animals and ultimately intelligent life (Catling *et al.*, 2005; Reinhard *et al.*, 2016). At every step of this progression, it is not clear to what degree evolutionary contingency has played a role and which developments would be inevitable given sufficient time and the appropriate environmental conditions. As a result, the capacity for evolution to generate diverse phototrophic and photosynthetic systems on Earth, even those that do not dominate today, may have considerable implications for the development of novel pigments on other habitable worlds (Johnson *et al.*, 2013).

### Retinal-based phototrophy as an astronomical biosignature

Regardless of the evolutionary sequence of events leading to retinal phototrophy on Earth, analog photopigments may have arisen independently in other habitable environments in the universe. For example, exoplanets within the habitable zones of most stars would receive ample photon fluxes to power significant levels of (bacterio)chlorophyll or rhodopsin analog phototrophy with some differences in total capacity based on the photospheric temperature and consequent spectral energy distribution of those stars (Kiang *et al.*, 2007b; 2007a; Komatsu *et al.*, 2015; Ritchie *et al.*, 2018). One exception may be the dimmest and reddest M-stars, which produce the least flux in the 400–700 nm wavelength range (Fig. 2b). For these stellar systems, total global productivity may be photon-limited rather than reductant or nutrient limited as it is on Earth (Lehmer *et al.*, 2018). However, FGK stellar systems are the more likely targets for future space-based direct-imaging missions capable of detecting astronomical biosignatures. This is due to the wider angular separation of star and planet in the habitable zone (Stark *et al.*, 2015; 2014) and the productivity of biospheres on planets orbiting these stars would not be photon-limited. It is, therefore, worthwhile to examine what the remote signatures of rhodopsin-like phototrophy would be on exoplanets and how they would compare to those produced by analogs to chlorophyll-based photosynthesis.

The most commonly referenced surface signature of life is the vegetation red edge (VRE), the steep increase in reflectivity of vegetation (primarily green vascular plants) at ∼700 nm (Gates *et al.*, 1965; Knipling, 1970). This increase is due to the contrast between the absorption of chlorophyll at red wavelengths and its high albedo at infrared wavelengths due to intracellular scattering. The VRE effect is commonly used to map vegetation by employing broadband observations from Earth-observing satellites (Huete *et al.*, 1994; Tucker *et al.*, 2005). While the VRE has been extensively examined as a possible exoplanet biosignature (Sagan *et al.*, 1993; Arnold *et al.*, 2002; Des Marais *et al.*, 2002; Seager *et al.*, 2005; Brandt and Spiegel, 2014), its applicability is limited to those planets that have, like Earth, evolved chlorophyll-analog (i.e., red-absorbing, infrared reflecting) powered vegetation with significant continental surface covering fractions.

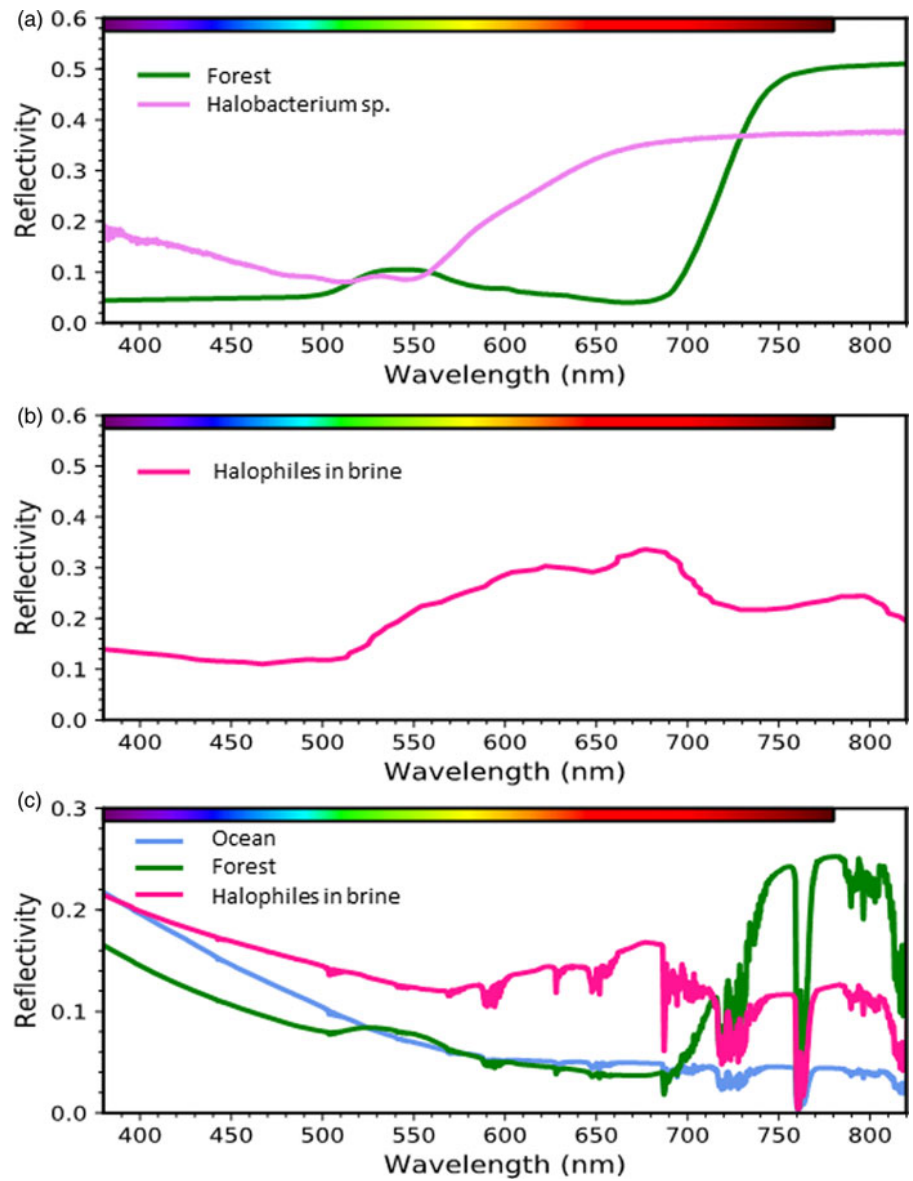

**Fig. 6.** Surface signatures of retinal and chlorophyll-based phototrophy. (a) Reflectance spectrum of a conifer forest (Baldridge *et al.*, 2009) and a culture of the phototrophic archaeon *Halobacterium* sp. (Schwieterman *et al.*, 2015). (b) Environmental spectrum of a halophile-dominated saltern pond in San Francisco Bay (Dalton *et al.*, 2009). (c) Simulated spectra of planets consisting of 100% sterile ocean, conifer forest, or a halophile-dominated saltern pond under an Earth-like atmosphere generated with a radiative transfer model (Schwieterman *et al.*, 2015).

Even on Earth, green vascular planets have only existed for just the last ~10% of the planet's history, about 470 million years out of 4.6-billion years (Kenrick and Crane, 1997). In order to address the limited time of presence of the VRE, astrobiologists interested in remote biosignatures have begun to consider and catalog surface reflectance signatures from a diverse array of known pigmented organisms including oxygenic and anoxygenic photosynthesizers, rhodopsin-based phototrophs and non-photosynthetic microbes that use pigments as a UV screen or antioxidant, or for other purposes (DasSarma, 2006; Kiang *et al.*, 2007b, 2007a; Cockell, 2014; Hegde *et al.*, 2015; Poch *et al.*, 2017; Schwieterman, 2018; Schwieterman *et al.*, 2018; 2015). The possible existence of a Purple Earth extends and expands the possible biological history of a planet when alternate biosignatures may be detectable, and it also enhances the number of possible evolutionary trajectories for which surface biosignatures may be found.

The photochemical properties of known prokaryotic rhodopsins on Earth are particularly worthy of study as a potential remote biosignature because of their capacity to generate chemical energy using an energy-rich portion of the electromagnetic spectrum. The most consequential difference between rhodopsin and chlorophyll-based phototrophy is the wavelength of maximum absorption. While the absorption peak of chlorophyll a is near 700 nm, bacteriorhodopsin absorption peaks near 570 nm. However, the expression of this signal would differ depending on whether the phototrophic organisms were on dry land or suspended in aquatic environments.

A true bacteriorhodopsin-based analog to terrestrial vegetation would possess a 'green-edge' comparable with the vegetation 'red-edge.' Fig. 6a illustrates the differences between the reflective spectra of a red-edge producing conifer forest (Baldridge *et al.*, 2009) and the green-edge producing Haloarchaea (Schwieterman *et al.*, 2015). While the green colour of the conifer forest results from inefficient absorption in a broad wavelength region centred near 550 nm, the pink colour of halophiles results from their high reflectivity at orange and red wavelengths due both to bacteriorhodopsin and carotenoid pigments such as bacterioruberins (Kushwaha and Kates, 1979; Oren *et al.*, 1992; Oren and Dubinsky, 1994). Green plants are similarly bright in the infrared due to their high reflectivity on the non-visible, long-wavelength side of the VRE. Notably, a wide variety of other organisms have spectral 'edges' at various visible wavelengths. Hegde *et al.* (2015) conducted an extensive series of reflectance measurements of plated cultures of extremophiles from 350 to 2500 nm showing a distribution of 'edge' features for various phototrophic and non-phototrophic species throughout the UV-visible spectrum. The radiotolerant species *Deinococcus radiodurans* also possess a 'green-edge' in plated cultures due to its primary pigment deinoxanthin (Cockell, 2014; Schwieterman *et al.*, 2015), which likely functions as an antioxidant. Depending on the environmental context, some of these pigments may also serve as alternative biosignatures. Rhodopsin-based (and other) phototrophs may have both biological and detectability advantage, however, in that they can harvest light energy for growth and accumulate at the surface of aquatic environments. Consequently, it is also important to consider the spectral differences between plated and suspended cultures, which would map to different planetary environments (e.g., land versus ocean).

The spectral signature of pigmented organisms like Haloarchea suspended in a lake or ocean would also be affected by the low reflectivity and strong absorption properties of aquatic environments. For example, halophilic communities present in saltern ponds possess a peak in brightness near ∼680 nm due to increasing reflectivity of bacteriorhodopsin and carotenoid pigments from the green to the red combined with strong water absorption at the reddest wavelengths (Fig. 6b; also see Dalton *et al.*, (2009)). Importantly, the brightness of halophilic pigments at orange and red wavelengths confers a detectability advantage over chlorophyll-containing cyanobacteria and algae suspended in water, because chlorophyll's high infrared reflectivity is counteracted by water vapour absorption, while chlorophyll is most absorptive at wavelengths where water is relatively transparent.

The remote signatures of these phototrophic organisms would further change from the spectral impact (i.e., absorption and scattering) of the overlying atmosphere (Fig. 6c). The spectral signatures of halophilic organisms are again somewhat favoured in this case because of the impact of overlying water vapour absorption nearly coincident with the VRE. Of course, the detectability of these signatures on an exoplanet will also be strongly sensitive to the land covering fraction, cell densities if suspended in water or brine and cloud cover effects (Sanromá *et al.*, 2014; Schwieterman *et al.*, 2015). The detectability potential of retinal photopigments and other halophilic pigment-analogs should be considered when anticipating the variety of potential surface biosignatures of exoplanets.

The capacity to detect surface biosignatures is an ongoing consideration in the design mandate of large, space-based telescopes (Fujii *et al.*, 2018; Schwieterman *et al.*, 2018) such as the conceived HabEx and LUVOIR/HDST missions (Dalcanton *et al.*, 2015; Mennesson *et al.*, 2016; Rauscher *et al.*, 2016; Bolcar *et al.*, 2017). Direct imaging spectra and spectrophotometry will allow characterization of the surfaces of terrestrial planets in the habitable zone, producing constraints on surface types, including surface biosignatures, providing the cloud covering fraction is sufficiently low (Sanromá *et al.*, 2014, 2013). The detectability potential of retinal photopigments and halophile-analogs suggests wavelengths shortward of the traditional VRE ($\lambda < 700$ nm) will be important to observe and analyse for 'edge' features suggestive of diverse phototrophic pigments and should be considered in the search for life outside the solar system.

## Concluding remarks

Although of enormous scientific interest, our understanding of the early evolutionary history of phototrophic life on Earth has remained limited. We propose here that the biochemical simplicity of retinal-based phototrophy, the spectral complementarity of bacteriorhodopsin pigments with chlorophylls and the newly uncovered widespread diversity of microbial rhodopsins throughout aquatic and terrestrial ecosystems are suggestive of the fundamental role retinal may have played in the early history of life on Earth. We posit here that domination by retinal-based phototrophs in the early history of life may have created the first 'Purple Earth' that at some point gave way to modern photosynthesizers before the rise of atmospheric oxygen. If correct, this early phototrophic metabolism would have greatly shaped the evolution of photosynthesis and indeed much of life on Earth. In fact, we know it continues to play a significant role in many environments today.

To test this Purple Earth hypothesis, future work should further explore natural communities of retinal-based phototrophs in diverse environments (e.g. arid, high altitude and polar locations). Additional studies are needed to explore the diversity and light capture capacity of retinal-based phototrophy in modern environments as they may continue to reveal unexpected roles and niches for this metabolism and inform its evolutionary origin. Additionally, future genomic analyses should be designed to consider the importance of the timing of the introduction of aerobic respiration in Haloarchaea in relation to the development of phototrophy during their metamorphosis from anaerobic chemolithoautotrophic methanogens to aerobic photoheterotrophs.

Considering an even broader view, the quest to understand the origin of life and early evolutionary events on our planet has gained increasing urgency with the discovery of thousands of new extra-solar planets, many of which are within the habitable zones of their host stars. Consequently, we may soon have the ability to characterize other potentially living worlds and finally answer the age-old question '*Are we alone in the universe?*' To realize this goal, however, we need to improve our understanding of major events sparking life on Earth and determine what biosignatures early life produced, especially those which may be detectable by remote sensing.

Simple retinal-based light-harvesting systems like that of the purple chromoprotein bacteriorhodopsin, may potentially serve as remote biosignatures for exoplanet research through the search for brightness peaks about 680 nm like that seen in hypersaline environments on Earth or by spectral 'edges' at green-yellow wavelengths (∼550 nm) analogous to the traditional vegetation 'red-edge' seen at 700 nm. These features are within the wavelength sensitivity window of planned next-generation space-based telescopes capable of directly imaging exoplanets and should be considered in the search for life in the universe.

**Acknowledgements.** Exobiology research in the S.D. laboratory is supported by NASA grant NNX15AM07G. E.S. is supported by a NASA Postdoctoral Fellowship, administered by the Universities Space Research Association and by the NASA Astrobiology Institute's Alternative Earths and Virtual Planetary Laboratory teams under Cooperative Agreement Nos. NNA15BB03A and NNA13AA93A, respectively. We thank Priya DasSarma for critical reading of the manuscript.

**References**

Abramov O and Mojzsis SJ (2009) Microbial habitability of the Hadean Earth during the late heavy bombardment. *Nature* **459**, 419–422.

Aouad M, Taib N, Oudart A, Lecocq M, Gouy M and Brochier-Armanet C (2018) Extreme halophilic archaea derive from two distinct methanogen class II lineages. *Molecular Phylogenetics and Evolution* **127**, 46–54.

Arnold L, Gillet S, Lardiere O, Riaud P and Schneider J (2002) A test for the search for life on extrasolar planets: looking for the terrestrial vegetation signature in the Earthshine spectrum. *Astronomy and Astrophysics* **237**, 7.

Atamna-Ismaeel N, Finkel OM, Glaser F, Sharon I, Schneider R, Post AF, Spudich JL, von Mering C, Vorholt JA, Iluz D, Béjà O and Belkin S (2012) Microbial rhodopsins on leaf surfaces of terrestrial plants. *Environmental Microbiology* **14**, 140–146.

Baldridge AM, Hook SJ, Grove CI and Rivera G (2009) The ASTER spectral library version 2.0. *Remote Sensing of Environment* **113**, 711–715.

Bayley H, Huang KS, Radhakrishnan R, Ross AH, Takagaki Y and Khorana HG (1981) Site of attachment of retinal in bacteriorhodopsin. *Proceedings of the National Academy of Sciences* **78**, 2225–2229.

Béjà O, Spudich EN, Spudich JL, Leclerc M and DeLong EF (2001) Proteorhodopsin phototrophy in the ocean. *Nature* **411**, 786–789.

Bell EA, Boehnke P, Harrison TM and Mao WL (2015) Potentially biogenic carbon preserved in a 4.1 billion-year-old zircon. *Proceedings of the National Academy of Sciences* **112**, 14518–14521.

Bolcar MR, Aloezos S, Crooke J, Dressing CD, Fantano L, Hylan JE, Tompkins S, Bolcar MR, Bly VT, Collins C, Feinberg LD, France K, Gochar G, Gong Q, Jones A, Linares I, Postman M, Pueyo L, Roberge A, Sacks L and West G (2017) The large UV/optical/infrared (LUVOIR) surveyor: decadal mission concept design update, In MacEwen H.A., Breckinridge J.B. (eds), *UV/Optical/IR Space Telescopes and Instruments: Innovative Technologies and Concepts VIII*. San Diego, CA: SPIE, pp. 9. doi:10.1117/12.2273848

Borda MJ, Elsetinow AR, Schoonen MA and Strongin DR (2001) *Astrobiology* **1**, 283–288.

Boucher Y and Doolittle WF (2000) The role of lateral gene transfer in the evolution of isoprenoid biosynthesis pathways. *Molecular Microbiology* **37**, 703–716.

Brandt TD and Spiegel DS (2014) Prospects for detecting oxygen, water, and chlorophyll on an exo-Earth. *Proceedings of the National Academy of Sciences* **111**, 13278–13283.

Brochier-Armanet C, Forterre P and Gribaldo S (2011) Phylogeny and evolution of the Archaea: one hundred genomes later. *Current Opinion in Microbiology* **14**, 274–281.

Brocks JJ, Love GD, Summons RE, Knoll AH, Logan GA and Bowden SA (2005) Biomarker evidence for green and purple sulphur bacteria in a stratified Palaeoproterozoic sea. *Nature* **437**, 866–870.

Brown LS (2014) Proton-Pumping microbial rhodopsins – ubiquitous structurally simple helpers of respiration and photosynthesis. In Hohmann-Marriott MF (ed.), *The Structural Basis of Biological Energy Generation*. Dordrecht, Netherlands: Springer, pp. 1–20. doi:10.1007/978-94-017-8742-0_1

Buick R (2008) When did oxygenic photosynthesis evolve? . *Royal Society of London Philosophical Transactions Series B* **363**, 2731–2743.

Catling DC, Glein CR, Zahnle KJ and McKay CP (2005) Why $O_2$ is required by complex life on habitable planets and the concept of planetary "oxygenation time". *Astrobiology* **5**, 415–438.

Chen M and Blankenship RE (2011) Expanding the solar spectrum used by photosynthesis. *Trends in Plant Science* **16**, 427–431.

Chen M, Schliep M, Willows RD, Cai Z-L, Neilan BA and Scheer H (2010) A red-shifted chlorophyll. *Science* **329**, 1318–1319.

Cockell CS (2014) Habitable worlds with no signs of life. *Philosophical Transactions. Series A, Mathematical, Physical, and Engineering Sciences* **372**, 20130082.

Cohen Y and Rosenberg E (1989) *Microbial mats: physiological ecology of benthic microbial Communities*.

Cohn CA, Mueller S, Wimmer E, Leifer N, Greenbaum S, Strongin DR and Schoonen AAA (2006) Pyrite-induced hydroxyl radical formation and its effect on nucleic acids. *Geochemical Transactions* **7**, 3.

Crick FH (1968) The origin of the genetic code. *Journal of Molecular Biology* **38**, 367–379.

Croce R and van Amerongen H (2014) Natural strategies for photosynthetic light harvesting. *Nature Chemical Biology* **10**, 492–501.

Dalcanton J, Seager S, Aigrain S, Hirata C, Battel S, Mather J, Brandt N, Postman M, Conroy C, Redding D, Feinberg L, Schiminiovich D, Gezari S, Stahl HP, Guyon O, Tumilinson J and Harris W (2015) From Cosmic Birth to Living Earths: The Future of UVOIR Space Astronomy. From Cosmic Births to Living Earths Report. (Washington, DC: Association for Research in Astronomy), Available at http://www.hdstvision.org/report/.

Dalton JB, Palmer-Moloney LJ, Rogoff D, Hlavka C and Duncan C (2009) Remote monitoring of hypersaline environments in San Francisco Bay, CA, USA. *International Journal of Remote Sensing* **30**, 2933–2949.

DasSarma S (2006) Extreme halophiles are models for astrobiology. *Microbe-American Society for Microbiology* **1**, 120–126.

DasSarma P, Zamora RC, Müller JA and DasSarma S (2012) Genome-wide responses of the model Archaeon Halobacterium sp. Strain NRC-1 to oxygen limitation. *Journal of Bacteriology* **194**, 5530–5537.

Deamer D (2011) *First Life: Discovering the Connections Between Stars, Cells, and how Life Began*. Oakland, CA: University of California Press.

Des Marais D (2000) When did photosynthesis emerge on Earth? *Science* **289**, 1703–1705.

Des Marais DJ, Harwit MO, Jucks KW, Kasting JF, Lin DNC, Lunine JI, Schneider J, Seager S, Traub WA and Woolf NJ (2002) Remote sensing of planetary properties and biosignatures on extrasolar terrestrial planets. *Astrobiology* **2**, 153–181.

Ernst OP, Lodowski DT, Elstner M, Hegemann P, Brown LS and Kandori H (2014) Microbial and animal rhodopsins: structures, functions, and molecular mechanisms. *Chemical Reviews* **114**, 126–163.

Fenchel T (2002) *Origin and Early Evolution of Life*. Oxford, UK: Oxford University Press.

Frigaard N, Larsen KL and Cox RP (1996) Spectrochromatography of photosynthetic pigments as a fingerprinting technique for microbial phototrophs. *FEMS Microbiology Ecology* **20**, 69–77.

Fujii Y, Angerhausen D, Deitrick R, Domagal-Goldman S, Grenfell JL, Hori Y, Kane SR, Pallé E, Rauer H, Siegler N, Stapelfeldt K and Stevenson KB (2018) Exoplanet biosignatures: observational prospects. *Astrobiology* **18**, 739–778.

Gates DM, Keegan HJ, Schleter JC and Weidner VR (1965) Spectral properties of plants. *Applied Optics* **4**, 11.

Goldsworthy A (1987) Why did nature select green plants? *Nature* **328**, 207–208.

Gómez-Consarnau L, Levine NM, Cutter LS, Wang D, Seegers B, Arístegui J, Fuhrman JA, Gasol JM and Sañudo-Wilhelmy SA (2017) Marine proteorhodopsins rival photosynthesis in solar energy capture. *Biorxiv* **231167**, 1–6. DOI: 10.1101/231167.

Guerrero LD, Vikram S, Makhalanyane TP and Cowan DA (2017) Evidence of microbial rhodopsins in Antarctic Dry Valley edaphic systems. *Environmental Microbiology* **19**, 3755–3767.

Gunner MR, Amin M, Zhu X and Lu J (2013) Molecular mechanisms for generating transmembrane proton gradients. *Biochimica et Biophysica Acta - Bioenergetics* **1827**, 892–913.

Hahn J and Haug P (1986) Traces of archaebacteria in ancient sediments. *Systematic and Applied Microbiology* **7**, 178–183.

Hallam SJ, Putnam N, Preston CM, Detter JC, Rokhsar D, Richardson PH and DeLong EF (2004) Reverse methanogenesis: testing the hypothesis with environmental genomics. *Science* **305**, 1457–1462.

Hedges SB (2002) The origin and evolution of model organisms. *Nature Reviews Genetics* **3**, 838–849.

Hegde S, Paulino-Lima IG, Kent R, Kaltenegger L and Rothschild L (2015) Surface biosignatures of exo-Earths: remote detection of extraterrestrial life. *Proceedings of the National Academy of Sciences* **112**, 3886–3891.

Henderson R and Unwin PNT (1975) Three-dimensional model of purple membrane obtained by electron microscopy. *Nature* **257**, 28–32.

Hirai T, Subramaniam S and Lanyi JK (2009) Structural snapshots of conformational changes in a seven-helix membrane protein: lessons from bacteriorhodopsin. *Current Opinion in Structural Biology* **19**, 433–439.

Hoshino Y and Gaucher EA (2018) On the origin of isoprenoid biosynthesis. *Molecular Biology and Evolution* **35**, 2185–2197.

Huete A, Justice C and Liu H (1994) Development of vegetation and soil indexes for MODIS-EOS. *Remote Sensing of Environment* **49**, 224–234.

Hylemon PB and Harder J (1998) Biotransformation of monoterpenes, bile acids, and other isoprenoids in anaerobic ecosystems. *FEMS Microbiology Reviews* **22**, 475–488.

Jeffrey SW (1963) Purification and properties of chlorophyll c from *Sargassum flavicans*. *Biochemical Journal* **86**, 313.

Johnson N, Zhao G, Caycedo F, Manrique P, Qi H, Rodriguez F and Quiroga L (2013) Extreme alien light allows survival of terrestrial bacteria. *Scientific Reports* **3**, 2198.

Kandori H (2015) Ion-pumping microbial rhodopsins. *Frontiers in Molecular Biosciences* **2**, 52.

Kennedy SP, Ng WV, Salzberg SL, Hood L and DasSarma S (2001) Understanding the adaptation of species NRC-1 to its extreme environment through computational analysis of its genome sequence. *Genome Research* **11**, 1641–1650.

Kenrick P and Crane PR (1997) The origin and early evolution of plants on land. *Nature* **389**, 33–39.

Kiang NY, Segura A, Tinetti G, Govindjee, Blankenship RE, Cohen M, Siefert J, Crisp D and Meadows VS, (2007a). Spectral signatures of photosynthesis. II. Coevolution with other stars and the atmosphere on extrasolar worlds. *Astrobiology* **7**, 252–274.

Kiang NY, Siefert J, Govindjee and Blankenship RE, (2007b). Spectral signatures of photosynthesis. I. Review of Earth organisms. *Astrobiology* **7**, 222–251.

Knipling EB (1970) Physical and physiological basis for the reflectance of visible and near-infrared radiation from vegetation. *Remote Sensing of Environment* **1**, 155–159.

Knoll AH (2015) *Life on A Young Planet: The First Three Billion Years of Evolution on Earth*. Princeton, NJ: Princeton University Press, Princeton. doi:10.1515/9781400866045

Komatsu Y, Umemura M, Shoji M, Kayanuma M, Yabana K and Shiraishi K (2015) Light absorption efficiencies of photosynthetic pigments: the dependence on spectral types of central stars. *International Journal of Astrobiology* **14**, 505–510.

Krebs MP and Khorana HG (1993) Mechanism of light-dependent proton translocation by bacteriorhodopsin. *Journal of Bacteriology* **175**, 1555–1560.

Kump LR (2008) The rise of atmospheric oxygen. *Nature* **451**, 277–278.

Kushwaha SC and Kates M (1979) Studies of the biosynthesis of C 50 carotenoids in *Halobacterium cutirubrum*. *Canadian Journal of Microbiology* **25**, 1292–1297.

Lange BM, Rujan T, Martin W and Croteau R (2000) Isoprenoid biosynthesis: the evolution of two ancient and distinct pathways across genomes. *Proceedings of the National Academy of Sciences USA* **97**, 13172–13177.

Laye VJ, Karan R, Kim J-M, Pecher WT, DasSarma P and DasSarma S (2017) Key amino acid residues conferring enhanced enzyme activity at cold temperatures in an Antarctic polyextremophilic β-galactosidase. *Proceedings of the National Academy of Sciences* **114**, 12530–12535.

Lehmer OR, Catling DC, Parenteau MN and Hoehler TM (2018) The productivity of oxygenic photosynthesis around cool, M dwarf stars. *The Astrophysical Journal* **859**, 171.

Luo G, Ono S, Beukes NJ, Wang DT, Xie S and Summons RE (2016) Rapid oxygenation of Earths atmosphere 2.33 billion years ago. *Science Advances* **2**, e1600134–e1600134.

Lyons TW, Reinhard CT and Planavsky NJ (2014) The rise of oxygen in Earth's early ocean and atmosphere. *Nature* **506**, 307–315. DOI: 10.1038/nature13068.

Mailloux RJ, Bériault R, Lemire J, Singh R, Chénier DR, Hamel RD and Appanna VD (2007) The Tricarboxylic acid cycle, an ancient metabolic network with a novel twist. *PLoS ONE* **2**, e690

Mancinelli RL and Hochstein LI (1986) The occurrence of denitrification in extremely halophilic bacteria. *FEMS Microbiology Letters* **35**, 55–58.

Mazor Y, Greenberg I, Toporik H, Béjà O and Nelson N (2012) The evolution of photosystem I in light of phage-encoded reaction centers. *Philosophical Transactions of the Royal Society B: Biological Sciences* **367**, 3400–3405.

Meadows VS, Arney GN, Schwieterman EW, Lustig-Yaeger J, Lincowski AP, Robinson T, Domagal-Goldman SD, Deitrick R, Barnes RK, Fleming DP, Luger R, Driscoll PE, Quinn TR and Crisp D (2018) The habitability of *Proxima centauri* b: environmental states and observational discriminants. *Astrobiology* **18**, 133–189.

Mennesson B, Gaudi S, Seager S, Cahoy K, Domagal-Goldman S, Feinberg L, Guyon O, Kasdin J, Marois C, Mawet D, Tamura M, Mouillet D, Prusti T, Quirrenbach A, Robinson T, Rogers L, Scowen P, Somerville R, Stapelfeldt K, Stern D, Still M, Turnbull M, Booth J, Kiessling A, Kuan G and Warfield K (2016) The Habitable Exoplanet (HabEx) imaging mission: preliminary science drivers and technical requirements. In MacEwen HA, Fazio GG, Lystrup M, Batalha N, Siegler N and Tong EC (eds), *Proc. SPIE 9904, Space Telescopes and Instrumentation 2016: Optical, Infrared, and Millimeter Wave*. Edinburgh, UK: Society of Photo-Optical Instrumentation Engineers (SPIE), pp. 99040L-1–99040L-10. doi:10.1117/12.2240457

Mitchell P (1961) Coupling of phosphorylation to electron and hydrogen transfer by a chemi-osmotic type of mechanism. *Nature* **191**, 144–148.

Müller JA and DasSarma S (2005) Genomic analysis of anaerobic respiration in the Archaeon Halobacterium sp. Strain NRC-1: dimethyl sulfoxide and trimethylamine N-oxide as terminal electron acceptors. *Journal of Bacteriology* **187**, 1659–1667.

Nelson-Sathi S, Dagan T, Landan G, Janssen A, Steel M, McInerney JO, Deppenmeier U and Martin WF (2012) Acquisition of 1,000 eubacterial genes physiologically transformed a methanogen at the origin of Haloarchaea. *Proceedings of the National Academy of Sciences* **109**, 20537–20542.

Nisbet EG, Grassineau NV, Howe CJ, Abell PI, Regelous M and Nisbet RER (2007) The age of Rubisco: the evolution of oxygenic photosynthesis. *Geobiology* **5**, 311–335.

Nutman AP, Bennett VC, Friend CRL, Van Kranendonk MJ and Chivas AR (2016) Rapid emergence of life shown by discovery of 3,700-million-year-old microbial structures. *Nature* **537**, 535–538.

Ohtomo Y, Kakegawa T, Ishida A, Nagase T and Rosing MT (2014) Evidence for biogenic graphite in early Archaean Isua metasedimentary rocks. *Nature Geoscience* **7**, 25–28.

Olson JM (2006) Photosynthesis in the Archean era. *Photosynthesis Research* **88**, 109–117.

Oren A and Dubinsky Z (1994) On the red coloration of saltern crystallizer ponds. II. Additional evidence for the contribution of halobacterial pigments. *International Journal of Salt Lake Research* **3**, 9–13.

Oren A, Stambler N and Dubinsky Z (1992) On the red coloration of saltern crystallizer ponds. *International Journal of Salt Lake Research* **1**, 77–89.

PBS Eons (2018) When the Earth was purple. Available at https://www.pbs.org/video/when-the-earth-was-purple-acpjlv/

Pinhassi J, DeLong EF, Béjà O, González JM and Pedrós-Alió C (2016) Marine bacterial and Archaeal Ion-pumping rhodopsins: genetic diversity, physiology, and ecology. *Microbiology and Molecular Biology Reviews* **80**, 929–954.

Planavsky NJ, Asael D, Hofmann A, Reinhard CT, Lalonde SV, Knudsen A, Wang X, Ossa Ossa F, Pecoits E, Smith AJB, Beukes NJ, Bekker A, Johnson TM, Konhauser KO, Lyons TW and Rouxel OJ (2014) Evidence for oxygenic photosynthesis half a billion years before the great oxidation event. *Nature Geoscience* **7**, 283–286.

Poch O, Frey J, Roditi I, Pommerol A, Jost B and Thomas N (2017) Remote sensing of potential biosignatures from rocky, liquid, or Icy (Exo)Planetary surfaces. *Astrobiology* **17**, 231–252.

Racker E and Stoeckenius W (1974) Reconstitution of purple membrane vesicles catalyzing light-driven proton uptake and adenosine triphosphate formation. *The Journal of Biological Chemistry* **249**, 662–663.

Rangarajan R, Galan JF, Whited G and Birge RR (2007) Mechanism of spectral tuning in green-absorbing proteorhodopsin. *Biochemistry* **46**, 12679–12686.

Rauscher BJ, Canavan ER, Moseley SH, Sadleir JE and Stevenson T (2016) Detectors and cooling technology for direct spectroscopic biosignature characterization. *Journal of Astronomical Telescopes, Instruments, and Systems* **2**, 041212.

Reinhard CT, Planavsky NJ, Olson SL, Lyons TW and Erwin DH (2016) Earth's oxygen cycle and the evolution of animal life. *Proceedings of the National Academy of Sciences* **113**, 8933–8938.

Ritchie RJ, Larkum AWD and Ribas I (2018) Could photosynthesis function on *Proxima centauri* b? *International Journal of Astrobiology* **17**, 147–176.

**Rothschild LJ** (2008) The evolution of photosynthesis…again? *Philosophical Transactions of the Royal Society B* **363**, 2787–2801.

**Sagan C, Thompson WR, Carlson R, Gurnett D and Hord C** (1993) A search for life on Earth from the Galileo spacecraft. *Nature* **365**, 715–721.

**Sanromá E, Pallé E and García Munõz A** (2013) On The effects of the evolution of microbial mats and land plants on the earth as a planet. Photometric and spectroscopic light curves of paleo-Earths. *The Astrophysical Journal* **766**, 133.

**Sanromá E, Pallé E, Parenteau MN, Kiang NY, Gutiérrez-Navarro aM, López R and Montañés-Rodríguez P** (2014) Characterizing the purple Earth: modeling the globally integrated spectral variability of the Archean Earth. *The Astrophysical Journal* **780**, 52.

**Schwieterman EW** (2018) Surface and temporal biosignatures. In Deeg H and Belmont J (eds), *Handbook of Exoplanets*. Cham, Switzerland: Springer International Publishing, pp. 1–29. DOI: 10.1007/978-3-319-30648-3_69-1.

**Schwieterman EW, Cockell CS and Meadows VS** (2015) Nonphotosynthetic pigments as potential biosignatures. *Astrobiology* **15**, 341–361.

**Schwieterman EW, Kiang NY, Parenteau MN, Harman CE, DasSarma S, Fisher TM, Arney GN, Hartnett HE, Reinhard CT, Olson SL, Meadows VS, Cockell CS, Walker SI, Grenfell JL, Hegde S, Rugheimer S, Hu R and Lyons TW** (2018) Exoplanet biosignatures: a review of remotely detectable signs of life. *Astrobiology* **18**, 663–708.

**Seager S, Turner EL, Schafer J and Ford EB** (2005) Vegetation's Red edge: a possible spectroscopic biosignature of extraterrestrial plants. *Astrobiology* **5**, 372–390.

**Segura A, Krelove K, Kasting JF, Sommerlatt D, Meadows V, Crisp D, Cohen M and Mlawer E** (2003) Ozone concentrations and ultraviolet fluxes on Earth-like planets around other stars. *Astrobiology* **3**, 689–708.

**Söll D and RajBhandary UL** (2006) The genetic code - thawing the "frozen accident". *Journal of Biosciences* **31**, 459–463.

**Stark CC, Roberge A, Mandell A and Robinson TD** (2014) Maximizing The exoearth candidate yield from a future direct imaging mission. *The Astrophysical Journal* **795**, 122.

**Stark CC, Roberge A, Mandell A, Clampin M, Domagal-goldman SD, Mcelwain MW and Stapelfeldt KR** (2015) Lower limits on aperture size for an exoearth detecting coronagraphic mission. *The Astrophysical Journal* **808**, 149.

**Stevenson A, Burkhardt J, Cockell CS, Cray JA, Dijksterhuis J, Fox-Powell M, Kee TP, Kminek G, McGenity TJ, Timmis KN, Timson DJ, Voytek MA, Westall F, Yakimov MM and Hallsworth JE** (2015) Multiplication of microbes below 0.690 water activity: implications for terrestrial and extraterrestrial life. *Environmental Microbiology* **17**, 257–277.

**Stoeckenius W, Lozier RH and Bogomolni RA** (1979) Bacteriorhodopsin and the purple membrane of halobacteria. *Biochimica et Biophysica Acta (BBA) - Reviews on Bioenergetics* **505**, 215–278.

**Strous M and Jetten MSM** (2004) Anaerobic oxidation of methane and ammonium. *Annual Review of Microbiology* **58**, 99–117.

**Sumper M, Reitmeier H and Oesterhelt D** (1976) Biosynthesis of the purple membrane of Halobacteria. *Angewandte Chemie International Edition in English* **15**, 187–194.

**Tucker CJ, Pinzon JE, Brown ME, Slayback DA, Pak EW, Mahoney R, Vermote EF and El Saleous N** (2005) An extended AVHRR 8-km NDVI dataset compatible with MODIS and SPOT vegetation NDVI data. *International Journal of Remote Sensing* **26**, 4485–4498.

**Vankranendonk M, Philipott P, Lepot K, Bodorkos S and Piranjno F** (2008) Geological setting of Earth's oldest fossils in the ca. 3.5Ga dresser formation, pilbara craton, Western Australia. *Precambrian Research* **167**, 93–124.

**Ventura GT, Kenig F, Reddy CM, Schieber J, Frysinger GS, Nelson RK, Dinel E, Gaines RB and Schaeffer P** (2007) Molecular evidence of late Archean archaea and the presence of a subsurface hydrothermal biosphere. *Proceedings of the National Academy of Sciences* **104**, 14260–14265.

**Walter MR, Buick R and Dunlop JSR** (1980) Stromatolites 3,400–3,500 Myr old from the North pole area, Western Australia. *Nature* **284**, 443–445.

**Williamson A, Conlan B, Hillier W and Wydrzynski T** (2011) The evolution of photosystem II: insights into the past and future. *Photosynthesis Research* **107**, 71–86.

**Woese CR** (2002) On the evolution of cells. *Proceedings of the National Academy of Sciences* **99**, 8742–8747.

**Xiong J, Inoue K and Bauer CE** (1998) Tracking molecular evolution of photosynthesis by characterization of a major photosynthesis gene cluster from Heliobacillus mobilis. *Proceedings of the National Academy of Sciences* **95**, 14851–14856.

**Zannoni D** (ed.) (2004) *Respiration in Archaea and Bacteria, Advances in Photosynthesis and Respiration*. Netherlands, Dordrecht: Springer. doi: 10.1007/978-1-4020-3163-2